\documentclass[english,keywords,amsmath,amssymb,nofootinbib]{revtex4}
\usepackage[T1]{fontenc}
\usepackage[latin1]{inputenc}
\usepackage{babel}%
\usepackage{graphicx}
\usepackage{color}
\usepackage{bm}
\usepackage{longtable}
\usepackage{amsmath}
\usepackage{amsfonts}
\usepackage{amssymb}
\usepackage{array}
\usepackage{epstopdf}
\usepackage{epsfig}

\begin{document}
\title{Which verification qubits perform best for secure communication in noisy channel?}
\author{Rishi Dutt Sharma$^{1}$}
\author{Kishore Thapliyal$^{2}$}
\author{Anirban Pathak$^{2,}$ \footnote{anirban.pathak@gmail.com}}
\author{Alok Kumar Pan$^{1}$}
\author{Asok De$^{1}$}

\affiliation{$^{1}$ National Institute Technology Patna, Ashok Rajhpath, Patna, Bihar 800005, India}
\affiliation{$^{2}$Jaypee Institute of Information Technology, A-10, Sector-62,
Noida, UP-201307, India}
\begin{abstract}
In secure quantum communication protocols, a set of single qubits prepared using 2 or more mutually unbiased bases or a set of $n$-qubit ($n\geq2$)  entangled  states of a particular form are usually used to form a verification string which is subsequently used to detect traces of eavesdropping. The qubits that form a verification string are  referred to as decoy qubits, and there exists a large set of different quantum states that can be used as decoy qubits. In the absence of noise, any choice of decoy qubits  provides equivalent security. In this paper, we examine such equivalence for noisy environment (e.g., in amplitude damping, phase damping, collective dephasing and collective rotation noise channels) by comparing the decoy-qubit assisted schemes of secure quantum communication that use single qubit states as decoy qubits with the schemes that use entangled states  as decoy qubits. Our study reveals that the single qubit assisted scheme perform better in some noisy environments, while some entangled qubits assisted schemes perform better in other noisy environments. Specifically, single qubits assisted schemes perform better in amplitude damping and phase damping noisy channels, whereas a few Bell-state-based decoy schemes are found to perform better in the presence of the collective noise. Thus, if the kind of noise present in a communication channel (i.e., the characteristics of the channel) is known or measured, then the present study can provide the best choice of decoy qubits required for implementation of schemes of secure quantum communication through that channel.
\end{abstract}
\maketitle
\section{Introduction}

In 1984, Bennett and Brassard first proposed an indigenous quantum key distribution (QKD) protocol \cite{bb84} (BB84 protocol) that can provide unconditional security, a feature which is considered to be unachievable in classical cryptography. Since this pioneering work of Bennett and Brassard,  the issue of unconditionally secure communication using the principles of quantum mechanics has been extensively studied both theoretically and experimentally \cite{bb84,ekert, b92,vaidman-goldenberg}.  Initial proposals for secure quantum communication were limited to QKD \cite{bb84,ekert, b92,vaidman-goldenberg}. However, later on several schemes for secure quantum communication tasks other than QKD have been proposed. For example, schemes were proposed for quantum secret sharing \cite{Hillery},  quantum key agreement \cite{QKA}, quantum secure direct communication (QSDC) \cite{Long-and-Liu,ping-pong,for-PP,lm05}, \textit{\emph{deterministic secure quantum communication}}\emph{ }(DSQC)
\cite{dsqc_summation,dsqcqithout-max-entanglement,dsqcwithteleporta,entanglement swapping,Hwang-Hwang-Tsai,reordering1,the:cao and song,the:high-capacity-wstate}. All these schemes of secure quantum communication can be broadly divided in two classes:

Class 1: Conjugate-coding-based schemes or BB84 type schemes, where 2 or more mutually unbiased bases (MUBs) are used to provide security. Specifically, these schemes provide security by utilizing the inability of the eavesdropper to perform simultaneous measurement in two or more MUBs. However, this is not a unique way to achieve the unconditional security, which can also be achieved using a single basis for both eavesdropping checking and encoding-decoding of information. This point was first noted by Goldenberg-Vaidman (GV) \cite{vaidman-goldenberg} who proposed first orthogonal-state-based protocol for QKD, which is now known as GV protocol, leading to a second class of communication protocols.

Class 2: Orthogonal-state-based protocols invoke a single basis for encoding, decoding and eavesdropping check.  An excellent example of orthogonal-state-based protocol is GV protocol. In this protocol, the security is obtained from duality, i.e., by making the special basis unavailable to Eve, so that, her measurements will leave a detectable trace \cite{With-preeti,With-chitra-ijtp,my-book}. Here, by special basis, we mean a basis set that includes  the initial quantum state or the state to be measured as a basis element (basis set used to prepare the initial state) as any measurement using this basis will lead to a deterministic result. Some of the present authors had shown in the recent past that the security of the orthogonal-state-based protocols with multipartite state arises due to monogamy of entanglement \cite{With-preeti,With-chitra-ijtp,my-book}. The  security of these schemes does not depend on conjugate coding and thus these protocols establish that conjugate coding is not essential for secure quantum communication. Due to this fact, these schemes are extremely appealing from the perspective of the foundational aspects of quantum mechanics.

We wish to note that in the schemes of both the classes, a set of qubits (that constitute the verification string) are measured by the authorized users to detect the presence of Eve. These qubits which are only used for eavesdropping check and thus cannot be employed for communication of a message and/or key distribution. Such qubits often referred to as decoy qubits or extra qubits as they are not directly used for communication of a message. Originally, the concept of decoy qubit (in context of QKD) was introduced with a slightly different purpose. In fact, it was introduced as a set of extra (in the sense that they are not used for communication of message or generation of keys) qubits which are intentionally prepared as multi-photon pulses and randomly mixed with single photon pulses (some of which will be used for communication or key generation) to differentiate between eavesdropping and channel noise and to circumvent photon number splitting (PNS) attack \cite{PNS}. Such decoy-qubit-based schemes of secure quantum communication have been experimentally realized by  various groups \cite{exp5,exp4,exp3,exp2,exp1}. Specifically, the first experimental realization was reported in 2006 \cite{exp5}. Subsequently, a long distance QKD scheme in optical fiber \cite{exp4} and free-space \cite{exp3} has also been implemented. Apart from weak coherent lights QKD schemes using polarization-based \cite{exp2} and parametric down conversion-based \cite{exp1} decoy states have also been experimentally implemented. A modified notion of decoy
qubits \cite{DLL,With-Anindita-pla,With-preeti,ap-cqsdc,my-book,With-chitra-IJQI,crypt-switch} was later used in DSQC and QSDC protocols, where the decoy qubits were viewed as (extra) qubits that were used only to detect the presence of Eve. This modified notion of decoy qubit is used in this work.

Various quantum states have been used to form verification strings for eavesdropping checking in schemes of secure quantum communication. For example, in BB84 protocol and in a large class of BB84-type protocols (such as  Ping-pong protocol in original form \cite{ping-pong}, LM05 protocol \cite{lm05}, DLL protocol \cite{DLL}, CL \cite{CL}, etc.),   a random sequence of single qubit states prepared in $\left\{ |0\rangle,\,|1\rangle\right\} $
and $\left\{ |+\rangle,\,|-\rangle\right\}$ bases is used as decoy qubits. If Eve measures all the travel qubits then she will choose half of the time wrong basis, which would lead to detectable (25 percent) errors if the receiver compares his measurement outcomes with the state prepared by the sender for those cases where  the same basis set is used by the sender and the receiver.  Recently, some of us were involved in a few works where we have established that the original GV protocol can be generalized to multipartite cases, and Bell states can be used as decoy qubits \cite{With-preeti,With-chitra-IJQI,my-book}. To be precise, $n$ copies of a specific bell state (say, $|\psi^{+}\rangle$) are used as decoy qubits and Alice (the sender) uses \textit{permutation of particles} (PoP) technique to spatially separate the entangled particles. This geographical separation makes the special basis unavailable to Eve. Consequently, an attempt for eavesdropping will leave enough detectable traces at the receiver's end. Specifically, an eavesdropping effort will cause entanglement swapping in the travel qubits, which will give other Bell states (other than $|\psi^{+}\rangle$) as well in the Bell measurement at the receiver's end \cite{ent-swap,ent-swap-sugata,ent-swap-with-chitra}. Further, entanglement swapping is not a characteristic of Bell states only, but can also be observed in all the entangled states. Recently, a similar strategy has been proposed using cluster state \cite{cluster-2012,Cluster-2015}, which the authors claimed to be
an improved scheme. As the security of the schemes that uses cluster state as decoy qubits is also achieved using entanglement swapping, the idea can be extended to use other multiqubit entangled states as verification qubits. In fact, all these eavesdropping checking techniques are equivalent, in the sense that they can be replaced by one another
without affecting the security of the protocol \cite{With-preeti,With-chitra-ijtp,my-book}. 

It is already established that to achieve the unconditional security for a decoy-qubit-based scheme, half of the qubits that travel through the channel accessible
to Eve are required to be checked for eavesdropping \cite{nielsen}. So, ideally, in QKD protocols, if the sender (Alice) wishes to make a key of $n$
bits she will have to use $2n$ qubits, out of which $n$ qubits are decoy qubits, i.e., $n$ qubits are required to be measured for eavesdropping checking. The number
of decoy qubits will further increase with the number of parties involved in a quantum key agreement protocol due to increase in the number
of communication channels. Thus, in secure quantum communication protocols, the same number of decoy qubits are used as the number
of message encoded qubits in each step of quantum communication.

In the ideal conditions, in  QKD and other schemes of secure quantum communication protocols, if the calculated error rate in eavesdropping checking step is below a tolerable limit, the parties taking part in the protocol proceed to the next step. Otherwise, they discard the protocol and start afresh. Hence, the decoy qubits play an important role in secure quantum communication. It is known that in the ideal situation, when communication channel is \textit{not noisy, any set of allowed decoy qubits provides an equivalent amount of security}. However, in a practical situation, the decoy qubits may interact with the environment, which may lead to decoherence, and thus the decoy qubits obtained at the receiver's end may not be exactly what were expected in the absence of noise. In such a scenario, it will be difficult to distinguish between the disturbance induced due to eavesdropping and the noise present in the channel. It would then be very interesting to examine the aforementioned equivalence of security irrespective of the choice of the decoy qubits when the communication channel is noisy. This simply made the motivation of the present work.

In this paper, we study the effects of different kinds of noise models, such as, amplitude damping (AD), phase damping (PD), collective dephasing (CD) and collective rotation (CR) noise models, on  different kind of decoy qubits that are proposed to be used to form verification string for eavesdropping detection. This specific choice of the noise models  is reasonable as both amplitude damping and phase damping noise models are extremely relevant in quantum optics and quantum communication. Further, these two noise models have also been experimentally verified in the recent past \cite{exp-noise}. Moreover, the collective noise model considers a coherent effect of noise on the travel qubits if the time delay between the photons is smaller than the variation of noise. The collective noise models (i.e., CD and CR) are important in those situations where we can consider that all the decoy qubits almost simultaneously travel through a noisy channel \cite{col-prl}. Here, we have compared the effects of noise on decoy qubits in terms of the fidelity of the decoy qubits prepared at the sender's end and the decoy qubits obtained at the receiver's end by considering various noisy quantum channels through which the decoy qubits travel, if no eavesdropping has been attempted. 

This paper is organized as follows. In Section \ref{sec:Various-strategies}, we summarize the various decoy-qubit-based strategies adopted so far in the literature. We studied and compared the effect of various kinds of noises on the decoy qubits in Section \ref{sec:Noise-models} and plotted the  fidelities obtained for different choices of decoy qubits with the noise parameters. We summarize and conclude our results in Section \ref{sec:Conclusions}. 

\section{Various strategies for detection of eavesdropping using decoy qubits\label{sec:Various-strategies}}

The security in different quantum cryptographic protocols arises from information versus disturbance trade-off which, in turn, determines the tolerable error limit for the sender and receiver \cite{ap-cqsdc}. In this section, we briefly discuss three types of strategies (usually referred to as subroutines) for detection of eavesdropping. These strategies differ from each other mainly on the choice of  decoy qubits. Specifically, in what follows, we discuss (i) BB84 subroutine, (ii) GV subroutine, and (iii) cluster-state-based subroutine.

In BB84 subroutine,  the sender uses two or more MUBs (say, $\left\{ |0\rangle,\,|1\rangle\right\} $
and $\left\{ |+\rangle,\,|-\rangle\right\} $), i.e., uses conjugate
coding to prepare the decoy qubits randomly in $\left\{ |0\rangle,\,|1\rangle\right\} $ or $\left\{ |+\rangle,\,|-\rangle\right\} $
basis. After an authentic acknowledgment of the receipt of all the particles from the receiver, the sender announces the relevant positions
of the decoy qubits. Subsequently, the receiver measures all the decoy
qubits randomly in either of the basis they were prepared and finally
announces the basis chosen for the measurements and the obtained outcomes
along with the positions of the respective decoy qubits. The sender now
compares the initial state of the decoy qubits with the receiver's
 outcomes only in those cases where the receiver chooses
the same basis as the sender had chosen to prepare that particular
decoy qubit. Ideally, only in the absence of eavesdropping the outcomes of the sender and receiver should
match. Thus, by choosing two or more non-orthogonal states as decoy qubits, the disturbance induced due to Eve's measurements can
be detected. This form of security can also be attributed
to no-cloning theorem along with the indistinguishability of the two non-orthogonal states \cite{my-book,ap-cqsdc}.

In GV subroutine,  instead of MUBs used in BB84 subroutine, one of the Bell states is used as decoy qubits. This subroutine runs as follows; once the receiver acknowledges the receipt of both the message and the decoy qubits, the sender discloses the positions of the decoy qubits, so that,
the receiver can perform Bell measurements on them which can detect the presence of Eve. Precisely, the disturbance at the receiver's end will be detected due to entanglement swapping caused by Eve's measurement. This is because, Eve has no knowledge about which  two particles are entangled, and this ignorance may lead her to perform Bell measurement on the wrong pair of particles, causing entanglement swapping. Consequently, at the receiver's end,
Bell measurement on the right pair of entangled decoy qubits will
result in the Bell states other than the one prepared by the sender as
decoy qubits. Unlike BB84 subroutine, the security arises due
to geographical or temporal separation of individual particles, which
makes the special basis unavailable to Eve. It would be apt to mention here that the security based on the detection
of Eve due to entanglement swapping caused by her measurement can
also be achieved for multi-qubit entangled states. The security for the schemes involving two or multi-qubit entangled states as decoy qubits is obtained by the monogamy of quantum entanglement \cite{With-preeti,With-chitra-ijtp}.
However, the generation of  Bell states is relatively easier than multipartite entanglement, which has made Bell-state-based GV subroutine widely accepted. 

In the cluster-state-based subroutine, the cluster states are used as decoy qubits. In Refs.  \cite{cluster-2012,Cluster-2015}, it is claimed that the
cluster-state-based scheme for eavesdropping detection is more efficient  than others. The cluster state is a four qubit entangled state given
by 
\begin{equation}
\left|\psi\right\rangle _{{\rm cluster}}=\frac{1}{2}\left(\left|0000\right\rangle +\left|0011\right\rangle +\left|1100\right\rangle -\left|1111\right\rangle \right).\label{eq:cluster-state}
\end{equation}
It is  evident that the security of protocols that use cluster-state-based subroutine
rely on the same principle as is used  in GV subroutine. A nice technique to achieve this geographical
separation of the entangled particles is PoP introduced by Deng and Long \cite{PoP} in 2003. Since then
many PoP-based protocols have been proposed, where the security is
achieved using Bell or cluster states as decoy qubits (\cite{With-chitra-IJQI,With-chitra-ijtp,With-Anindita-pla,With-preeti,crypt-switch}
and references therein). In fact, any entangled state could have been
used in this technique to facilitate the security, but the generation
and maintenance of multi-qubit entangled states is certainly more
difficult than that of the Bell states or single qubit states. Further, the generation of single qubit states is easier than that of the Bell states. This fact is the prime reason for
the prominent use of the single-qubit-based
BB84 subroutine.

We have already mentioned that the security of the protocols for secure quantum communication remains unchanged even
if one switches between different subroutines - a fact,  that has been established in Refs. \cite{my-book,ap-cqsdc}, where different
versions of quantum communication protocols are discussed using both
BB84 and GV subroutine. It is further established that the qubit
efficiency remains unchanged even if we change the subroutine adopted
for security. 

Note that, in an ideal communication channel, the security in all the decoy qubit assisted schemes lies
in the fact that any attempt of eavesdropping will leave detectable trace at the receiver's end. However, the presence of noise can also
lead to a mismatch at the receiver's end. It is then of deep interest  to study the range of decoy qubit assisted protocols in the presence of noisy channels. This simply sets the  motivation to study systematically the effect of different noise models on the various quantum states that have been used to form verification strings  in various protocols  of quantum communication.

\section{The effect of various kind of noises on decoy qubits\label{sec:Noise-models}}

We now study the effect of the interaction of the decoy qubits with the environment using different noise models. For this purpose, we consider a few specific noise models, such as, the AD, PD channels  and two kinds of collective noise channels, viz., the CD and CR noise channels. In order to study the effect of noise on various strategies of decoy qubits, we employ a method recently adopted in the works of some of the present authors \cite{crypt-switch,CBRSP-our-paper}. Let us first summarize the method that was used in Refs. \cite{crypt-switch,CBRSP-our-paper} for various noise models. In order to this, let us consider the density matrix for the initial quantum state  $\rho=|\psi\rangle\langle\psi|,$ where $|\psi\rangle$ is an $n$ qubit pure quantum state. Now, to study the effect of AD and PD noises, the transformed density
matrix in the presence of AD or PD channels can be written as 
\begin{equation}
\rho_{k}=\sum_{i_{j}}E_{i_{1}}^{k}\otimes E_{i_{2}}^{k}\otimes\cdots\otimes E_{i_{j}}^{k}\otimes\cdots\otimes E_{i_{n}}^{k}\rho\left(E_{i_{1}}^{k}\otimes E_{i_{2}}^{k}\otimes\cdots\otimes E_{i_{j}}^{k}\otimes\cdots\otimes E_{i_{n}}^{k}\right)^{\dagger},\label{eq:noise-effected-density-matrix-1}
\end{equation}
where $E_{i_{j}}^{k}$ are the suitable Kraus operators for AD or PD channels which will be explicitly mentioned later and $k$ denotes $A$ or $P$ for AD or PD channels, respectively. However, while considering collective noise models, the evolution of the density matrix of an $n$ qubit pure quantum state $\rho=|\psi\rangle\langle\psi|$   can be expressed as
\begin{equation}
\rho_{k}=U_{i}^{\otimes n}\rho U_{i}^{\dagger\otimes n},
\label{eq:noise-effected-density-matrix-1}
\end{equation}
where the subscript $k$ denotes $D$ or $R$ for CD or CR noise channels, and $U_{i}$ is a $2\times2$ unitary matrix (which operates on a single qubit) for either CD or CR noise channels.

Since, in the absence of any noise and eavesdropping, the expected quantum state at the receiver's end is same as that at the sender's end, i.e., $\rho=|\psi\rangle\langle\psi|$, the effect of noise can be determined by comparing it with the quantum state $\rho_{k}$ obtained in the presence of noisy channels. For this comparison, we can use fidelity, which is defined as \cite{fidel} 
\begin{equation}
F=\langle\psi|\rho_{k}|\psi\rangle.\label{eq:fidelity}
\end{equation}
Here, it would be apt to mention that the expression of fidelity considered in Eq. (\ref{eq:fidelity}) is slightly different from the conventional expression. Conventionally, fidelity of two quantum states $\rho$
  and $\sigma$
  is defined as $F_{c}(\sigma,\rho)=Tr\left[\sqrt{\sigma^{\frac{1}{2}}\rho\sigma^{\frac{1}{2}}}\right].$ Here, we have introduced a subscript $c$ to distinguish the conventional definition of fidelity from the one that is used in this paper. 
Clearly, for an ideal channel, the value of the $F$ should be unity. In what follows, we shall examine the effect of different kinds of noises in the communication channels using fidelity. 

As mentioned earlier, in this paper, we consider the AD, PD channels  and two kinds of collective noise channels, viz., the CD and CR noise channels.  We know that BB84 subroutine
uses random strings of $\left\{ \left|0\right\rangle ,\left|1\right\rangle ,\left|+\right\rangle ,\left|-\right\rangle \right\} $
as decoy qubits, while the same task in GV and cluster-state-based subroutines is performed using one of the Bell states and four qubit cluster
state, respectively. So, we can easily infer that  we require at least four qubits in BB84 subroutine
and two Bell states in GV subroutine to compare the BB84 and GV subroutines with the four-qubit cluster-state-based subroutine. Further, as in the BB84 subroutine random
strings of $\left\{ \left|0\right\rangle ,\left|1\right\rangle ,\left|+\right\rangle ,\left|-\right\rangle \right\} $ are used, to compare it with the other two subroutines the average of all possible 256 cases is obtained. The fidelity for various types of decoy qubits, when they are subjected to different noise models, are obtained and summarized in Table \ref{tab:fidelity}. The detailed analysis of the results obtained is discussed in following subsections.

\begin{center}
\begin{table}
\begin{centering}
\begin{tabular}{|>{\centering}m{1.8cm}|>{\centering}m{1.4cm}|c|>{\centering}p{2.55cm}|>{\centering}p{2.5cm}|>{\centering}p{2.4cm}|}
\hline 
Type of  & Decoy  & \multicolumn{1}{c}{} & \multicolumn{1}{>{\centering}p{2.55cm}}{Fidelity } & \multicolumn{1}{>{\centering}p{2.5cm}}{} & \tabularnewline
\cline{3-6} 
subroutine & qubits & In AD channel ($F_{A}$) & In PD channel ($F_{P}$) & In CD channel ($F_{D}$) & In CR channel ($F_{R}$)\tabularnewline
\hline 
BB84 subroutine  & Average of $\left\{ \left|0\right\rangle ,\left|1\right\rangle ,\right.$
$\left.\left|+\right\rangle ,\left|-\right\rangle \right\} $ & $\frac{1}{256}(3+\sqrt{1-\eta_{A}}-\eta_{A})^{4}$ & $\frac{1}{256}\left(-4+\eta_{P}\right)^{4}$ & $\frac{1}{256}(3+\cos\phi)^{4}$ & $\cos^{8}\theta$\tabularnewline
\hline 
 & $\left|\psi^{+}\psi^{+}\right\rangle $ & $\frac{1}{4}\left(2-2\eta_{A}+\eta_{A}^{2}\right)^{2}$ &  & $\cos^{4}\phi$ & 1\tabularnewline
\cline{2-2} \cline{6-6} 
GV  & $\left|\psi^{-}\psi^{-}\right\rangle $ &  &  &  & $\cos^{4}2\theta$\tabularnewline
\cline{2-3} \cline{5-5} 
subroutine  & $\left|\phi^{+}\phi^{+}\right\rangle $ & $\left(-1+\eta_{A}\right)^{2}$ &  & 1 & \tabularnewline
\cline{2-2} \cline{6-6} 
 & $\left|\phi^{-}\phi^{-}\right\rangle $ &  & $\frac{1}{4}\left(2-2\eta_{P}+\eta_{P}^{2}\right)^{2}$ &  & 1\tabularnewline
\cline{1-3} \cline{5-6} 
Cluster-state-based subroutine  & $\left|\psi\right\rangle _{{\rm cluster}}$ in Eq. (\ref{eq:cluster-state}) & $\frac{1}{4}(4-8\eta_{A}+6\eta_{A}^{2}-2\eta_{A}^{3}+\eta_{A}^{4})$ &  & $\cos^{4}\phi$ & $\cos^{8}\theta$\tabularnewline
\hline 
\end{tabular}
\par\end{centering}

\caption{\label{tab:fidelity} The expressions of the fidelities in various noise channels for different subroutines adopted for security checking. Here, the subscript $J$ in the fidelity $F_{J}$ denotes the noisy channels corresponding to AD, PD, CD, and CR noises.}
\end{table}

\par\end{center}

\subsection{Effect of amplitude damping (AD) noise}

We first consider the effect of AD noise in the quantum state $\rho$. In the AD noise, a dissipative interaction between a system and its environment is considered. Specifically, the environment is considered as a vacuum bath, i.e., a bath at zero temperature without any squeezing, and the interaction causes  loss of energy (photon) \cite{nielsen,preskill,SGAD}. Such a noise is characterized by the following Kraus operators 
\cite{nielsen,preskill,SGAD} 
\begin{equation}
E_{0}^{A}=\left[\begin{array}{cc}
1 & 0\\
0 & \sqrt{1-\eta_{A}}
\end{array}\right],\,\,\,\,\,\,\,\,\,\,\,\,\,\,\, E_{1}^{A}=\left[\begin{array}{cc}
0 & \sqrt{\eta_{A}}\\
0 & 0
\end{array}\right],\label{eq:Krauss-amp-damping} 
\end{equation}
where $\eta_{A}$ ($0\leq\eta_{A}\leq1$) is the decoherence rate
and describes the probability of error due to AD channel. In quantum optics and quantum communication, effect of AD noise is investigated very frequently  (\cite{exp-noise,our-QDs,CBRSP-our-paper,crypt-switch,fidel,ent-sud-death,AD-prot,AD-prot-exp} and references therein). For example, recently the effect of AD channels is studied in the context of nonclassical behavior of spin systems with open quantum system effects \cite{our-QDs}, controlled secure and insecure quantum communication \cite{CBRSP-our-paper,crypt-switch,fidel}, entanglement sudden death \cite{ent-sud-death}, and protecting remote bipartite entanglement subjected to this noise \cite{AD-prot}. It is interesting to note that a spin chain acts as an AD channel for quantum communication through it \cite{sugato}. Further, a set of experiments related to AD noise has also been carried out \cite{exp-noise}. These works \cite{exp-noise,our-QDs,CBRSP-our-paper,crypt-switch,ent-sud-death,AD-prot,AD-prot-exp} have clearly established the importance of AD noise and justifies our effort to study the effect of AD noise on decoy qubits. A similar logic is applicable to the other noise models studied here.

The effect of AD noise on different types of decoy qubits is calculated here by using the fidelity formula given in Eq. (\ref{eq:fidelity}) and the 
computed analytic expressions of the fidelities that are summarized in Table \ref{tab:fidelity}. We find that while using GV subroutine the fidelity for same
parity states is the same, i.e., the fidelity remains the same whether one
uses $\left|\psi^{+}\right\rangle $ or $\left|\psi^{-}\right\rangle $
($\left|\phi^{+}\right\rangle $ or $\left|\phi^{-}\right\rangle $)
Bell states, where $\left|\psi^{\pm}\right\rangle=\frac{\left|00\right\rangle{\pm}\left|11\right\rangle}{\sqrt2} $ and $\left|\phi^{\pm}\right\rangle=\frac{\left|01\right\rangle{\pm}\left|10\right\rangle}{\sqrt2}.$ 

\begin{figure}[h]
{\rotatebox{0}{\resizebox{8.0cm}{6.0cm}{\includegraphics{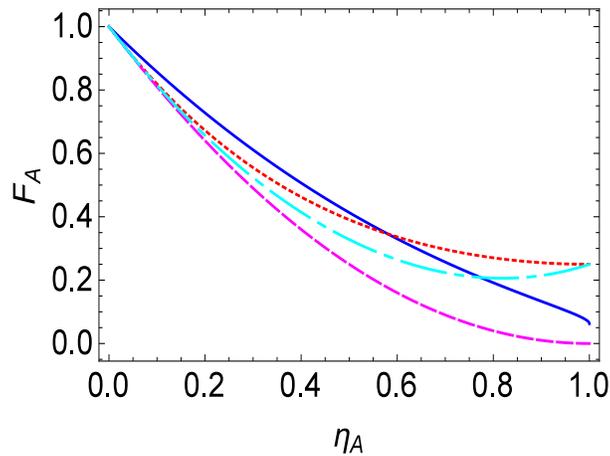}}}}
\caption{(Color online) The fidelity $F_{A}$ for various types of decoy qubits in the presence of AD noise are plotted with the decoherence rate $\eta_{A}$. The smooth (blue), dotted (red), dashed (magenta), and dot-dashed (cyan) lines correspond to the fidelities for BB84 subroutine, $\left|\psi^{\pm}\right\rangle $,
$\left|\phi^{\pm}\right\rangle $ and cluster state decoy-qubits-based subroutines, respectively.}
\label{fig:ad}
\end{figure}

The variation of the fidelities with decoherence rate ($\eta_{A}$) for various decoy qubits when subjected
to the AD noise are illustrated in Fig. \ref{fig:ad}. It can be observed from the figure  that for $\eta_{A}\leq0.5$, BB84 subroutine produces the maximum fidelity and hence is preferable. It can also be seen that fidelity for $\left|\psi^{\pm}\right\rangle $ ($\left|\phi^{\pm}\right\rangle $) decoy qubits is always greater (less) than the same for the cluster state. Following a similar approach, the interaction of the decoy qubits with a thermal bath, i.e., finite temperature bath, can also be investigated using generalized amplitude damping noise model \cite{nielsen,SGAD,our-QDs}, which can further be extended to the interaction with non-zero squeezing bath called squeezed thermal bath, and can be studied as decoy qubits exposed to squeezed generalized amplitude damping channel \cite{SGAD,our-QDs}.

\subsection{Effect of phase damping (PD) noise}

Let us now consider the effect of PD channel on different kinds of decoy qubits. In the PD noise, an interaction without loss of energy is considered between a system and its environment. Specifically, the effect of the environment is considered to vanish the off-diagonal terms of the density matrix, which leads to mixedness of the state. This is considered as a most natural kind of noise model \cite{nielsen,preskill} and is rigorously studied in contexts of various protocols of quantum communication  (\cite{CBRSP-our-paper,crypt-switch,fidel,ent-sud-death,err-det-exp-nmr} and references therein) and models of quantum optics (\cite{exp-noise,ent-sud-death,jay-cum} and references therein). The PD noise model is characterized by the following Kraus
operators \cite{nielsen,preskill}  
\begin{equation}
E_{0}^{P}=\sqrt{1-\eta_{P}}\left[\begin{array}{cc}
1 & 0\\
0 & 1
\end{array}\right],\,\,\,\,\,\,\,\,\,\,\,\,\,\,\, E_{1}^{P}=\sqrt{\eta_{P}}\left[\begin{array}{cc}
1 & 0\\
0 & 0
\end{array}\right],\,\,\,\,\,\,\,\,\,\,\,\,\,\,\, E_{2}^{P}=\sqrt{\eta_{P}}\left[\begin{array}{cc}
0 & 0\\
0 & 1
\end{array}\right],\label{eq:Krauss-phase-damping}
\end{equation}
where $\eta_{P}$ ($0\leq\eta_{P}\leq1$) is the decoherence rate
for the PD channels.

The effect of PD noise on different types of decoy qubits is computed by using the same strategy as was adopted for the AD noise. The 
analytic expressions of the fidelities that are computed here are listed in Table \ref{tab:fidelity}. Interestingly, it can be seen that the fidelities obtained 
 for GV and cluster-state-based subroutines are the same. This can be attributed to the fact that in the presence of PD noise, both $|0\rangle$ and $|1\rangle$ states are affected in the same manner (cf. matrix form of $E_{1}^{P}$ amd $E_{2}^{P}$). A similar nature can also be observed in the average fidelity with BB84 subroutine where the fidelity obtained for all the choices of states in both computational and diagonal basis are found to match exactly.

\begin{figure}[h]
{\rotatebox{0}{\resizebox{8.0cm}{6.0cm}{\includegraphics{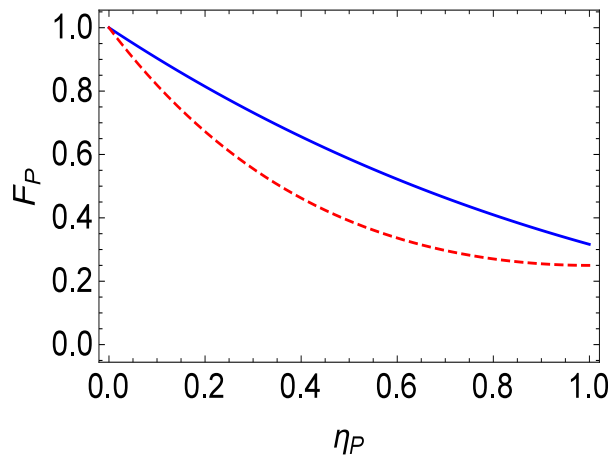}}}}
\caption{(Color online) The variation of fidelity $F_{P}$ for various types of decoy qubits in the presence of PD noise are plotted with the decoherence rate $\eta_{P}$. The smooth blue and dashed red lines correspond to the BB84 subroutine and the rest of the cases respectively.}
\label{fig:pd}
\end{figure}

The variation of the fidelity for different kinds of decoy qubits  with decoherence rate ($\eta_{P}$) are depicted in Fig. \ref{fig:pd}. 
It is observed that BB84 subroutine has larger fidelity than GV and cluster-state-based subroutines. Hence, BB84 subroutine is preferable in the PD noisy
environment. Thus, we may note that although entanglement is a costly resource it does not necessarily help, since in independent (non-collective) noise, the single qubits perfrom better than entangled decoys.

\subsection{Effect of collective dephasing (CD) noise }
We now consider the collective noise models to calculate the fidelity for various subroutines. Let us consider a scheme of secure quantum communication in which all the decoy qubits are transmitted almost simultaneously. In such a situation, the collective noise assumption \cite{col-prl} is satisfied. An arbitrary collective noise is considered as a situation in which all the qubits are coupled with the same environment. Interestingly, an arbitrary collective noise channel has been shown advantageous for entanglement distribution \cite{col-adv}. Further, it is also shown that the singlet state is decoherence free in an arbitrary collective noise \cite{col-Bell}. Let us first consider a CD noise model which is characterized by \cite{coll-noise} 
\begin{equation}
U_{p}\left|0\right\rangle =\left|0\right\rangle ,\,\,\,\,\,\,\,\,\,\,\,\,\,\, U_{p}\left|1\right\rangle =\exp\left(i\phi\right)\left|1\right\rangle ,\label{eq:coll-dephas}
\end{equation}
where $U_{p}$ is just a phase gate given by $\left[\begin{array}{cc}
1 & 0\\
0 & \exp\left(i\phi\right)
\end{array}\right]$ and $\phi$ is the noise parameter that can change with time but is same at an instant for all the qubits traveling through a noisy channel.

In CD noise, the parity-1 Bell states ($\left|\phi^{\pm}\right\rangle $)
are decoherence free as decoy qubits (cf. Table \ref{tab:fidelity}). 
Interestingly, this result is consistent with the results reported
by Li \textit{et al. }  \cite{coll-noise}, where the authors claimed that these anti-parallel Bell states
form a decoherence free subspace under this particular noise. This fact establishes that the parity-1 Bell states are best choices of decoy qubits with GV subroutine in CD noisy environment\footnote{Except the parity-1 Bell states, W state is also observed to be decoherence free in CD noise. Here, we restrict our discussion only up to Bell states. However, we note that W state is also found to be an excellent choice as decoy qubits for a channel with CD noise.}. As collective noise is one of the predominant causes
of decoherence in the experimental implementation of quantum communication, it is important to find decoherence-free states, which can protect quantum information from collective noise. Interestingly, decoherence free nature of a four qubit  quantum state used for encoding of a qubit has been experimentally verified by Bourennane et al. \cite{col-exp} using quantum state tomography. Similar techniques may be adopted to verify the findings of the present work.
\begin{figure}[h]
{\rotatebox{0}{\resizebox{8.0cm}{6.0cm}{\includegraphics{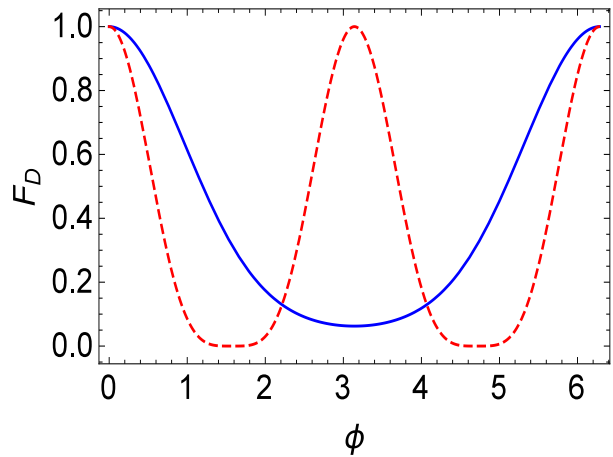}}}}
\caption{(Color online) The dependence of the fidelity $F_{D}$ on the CD noise parameter, i.e., the phase angle $\phi$, is illustrated for various types of decoy qubits exposed to CD noise channel. The smooth blue curve and the dashed red line correspond the BB84 subroutine  and  $\left|\psi^{\pm}\right\rangle $
or cluster state respectively.}
\label{fig:cd}
\end{figure}

For the remaining choices of decoy qubits, the variation of fidelity
with phase angle is shown in Fig. \ref{fig:cd}. Interestingly,
we can observe that when $\left|\psi^{\pm}\right\rangle $ are taken as decoy
qubits (as in GV subroutine)  and in cluster-state-based subroutine, the fidelity becomes
unity for phase angle $\phi=n\pi/2$, and it becomes zero
for $\phi=\left(2n+1\right)\pi/2$. However, the average fidelity
in BB84 subroutine do not show this kind of behavior.

\subsection{Effect of collective rotation (CR) noise }

Now, we consider another collective noise model: the CR noise model, which can be characterized as \cite{coll-noise} 
\begin{equation}
U_{r}\left|0\right\rangle =\cos\theta\left|0\right\rangle +\sin\theta\left|1\right\rangle ,\,\,\,\,\,\,\,\,\,\,\,\,\,\, U_{r}\left|1\right\rangle =-\sin\theta\left|0\right\rangle +\cos\theta\left|1\right\rangle ,\label{eq:coll-rotation}
\end{equation}
where $U_{r}$ is a unitary rotation given by $\left[\begin{array}{cc}
\cos\theta & -\sin\theta\\
\sin\theta & \cos\theta
\end{array}\right]$ and  $\theta$ is the noise parameter that fluctuates with time, similar to $\phi$ in the case of CD noise, but remains the same for all the qubits traveling simultaneously through a noisy channel.

\begin{figure}[h]
{\rotatebox{0}{\resizebox{8.0cm}{6.0cm}{\includegraphics{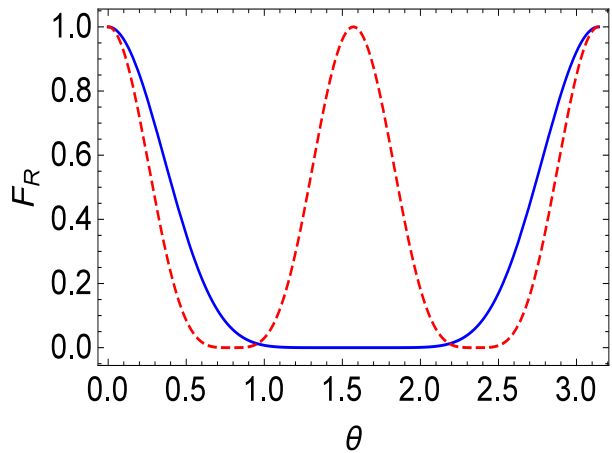}}}}
\caption{(Color online) The variation of fidelity
with rotation angle $\theta$ has been illustrated, when decoy qubits are subjected to
CR noise. The smooth (blue) line represents the BB84 and cluster-state-based
subroutine, while the dashed (red) line corresponds to the GV subroutine
with $\left|\psi^{-}\right\rangle $ and $\left|\phi^{+}\right\rangle $
decoy qubits.}
\label{fig:cr}
\end{figure}

The calculated fidelity expressions for various decoy qubits when subjected to CR noise channels
are summarized in Table \ref{tab:fidelity}. It can be seen that $\left|\psi^{+}\right\rangle $ and $\left|\phi^{-}\right\rangle $
states are decoherence free. This is consistent with the results of Ref. \cite{coll-noise}.
A detailed analysis of the remaining cases reveals that the average fidelity in BB84 subroutine
matches exactly with the fidelity obtained using cluster states. Similarly,
the fidelity of $\left|\psi^{-}\right\rangle $ and $\left|\phi^{+}\right\rangle $
states are the same. These two expressions of fidelities are graphically shown in Fig. \ref{fig:cr},
where we can see that for $\theta\in\left[\frac{\pi}{3},\frac{2\pi}{3}\right]$
average fidelity in BB84 subroutine approaches to zero, and thus GV subroutine
is preferable with $\left|\psi^{-}\right\rangle $ and $\left|\phi^{+}\right\rangle $
states as decoy qubits. In contrary, the BB84 subroutine yields higher
fidelity in the regions beyond this particular region, i.e., $\theta\leq\frac{\pi}{3}$
and $\theta\geq\frac{2\pi}{3}$. However, our investigation reveals that GV subroutine with $\left|\psi^{+}\right\rangle $ or $\left|\phi^{-}\right\rangle $ as the decoy qubits provide us the best choice for decoy qubits for a channel with CR noise. This is so as the states $\left|\psi^{+}\right\rangle $ and $\left|\phi^{-}\right\rangle $ are decoherence free in CR noise.

\section{Conclusions\label{sec:Conclusions}}
 It is known that in  the absence of noise in the communication channels, all the decoy-qubit-based techniques available for
detection of eavesdropping are equivalent. However, in the realistic
situations various types of noises present in the communication channel may challenge this equivalence and thus lead to a preferred decoy state for a specific type of noise. Keeping that in mind, in this paper, we have investigated the effect of various kinds of noise models on a set of  quantum states that can be used as decoy qubits. Specifically, here, we have  developed a clear strategy for comparison of securities provided by BB84, GV and cluster-state-based subroutines in various types of noisy environments. The investigation performed using our strategy has yielded several interesting results. For example, we have observed that the single qubit decoy states used in BB84 subroutine usually perform better in noisy environments if the nature of interaction between the decoy qubits and the surrounding is not known. However, the same is not true in general (i.e., if the suitable noise model has been characterized). Specifically, in the presence of AD noise, the BB84 subroutine is observed to yield largest fidelity up to moderate decoherence rate ($\eta_{A}<0.6$), while $\left|\psi^{\pm}\right\rangle$ performs better when the decoherence rate becomes very high and get saturated to a value $F_{A}=0.3$. In PD noise model as well, the  BB84 subroutine is found to perform much 
better than the other decoy-qubit-based subroutines implemented using various entangled states. In fact, all the other schemes investigated here are observed to provide the same fidelity in the PD channel. 
Similarly, for all the cases, where fidelity in the presence of CD noise depends on the noise parameter (phase angle), all the other schemes except BB84 subroutine are found to provide the same fidelity. But, no protocol is found to be advantageous in comparison to the other as the expressions of fidelity are found to be funtion of the parameter $\phi$, unlike $\left|\phi^{\pm}\right\rangle$ which is a decoherence free state in a CD channel. For example, for $\phi=\frac{(2n+1)}{2}\pi$, BB84 subroutine is found to perform relatively better than the remaining cases, but for $\phi=n\pi$ all other subroutines provide maximum fidelity. A similar study in context of CR noise model reveals that GV subroutine (when fidelity is rotation angle dependent) have maximum fidelity for $\theta=\frac{(2n+1)}{2}\pi$ whereas both BB84 and cluster-state-based subroutines provide ideal results only for $\theta=n\pi$.

In the above, we have observed that an appropriate choice of quantum state to
be used as decoy qubit depends on the character of the channel (i.e., on which noise is present in the system).
This point is illustrated clearly in the context of GV subroutine, where we have seen that in the presence of AD noise worst choice for decoy qubits is the states  $\left|\phi^{\pm}\right\rangle$. However, $\left|\phi^{-}\right\rangle$ provides us the best choice for decoy qubits in a noisy channel with CD noise, CR noise or both the noises. To recognize this interesting fact, we may note that we have observed that in the presence of CD noise, a decoherence free subspace is formed by the quantum states 
$\left\{\left|\phi^{\pm}\right\rangle \right\}$. Thus, in presence of CD noise, it is clearly beneficial to use one of  the states from the set $\left\{\left|\phi^{\pm}\right\rangle \right\}$  as decoy qubits. If Alice and Bob use one of these Bell states as decoy qubits and the receiver
is found to  obtain another Bell state on his/her Bell measurement
on the partner particles, that would certainly imply the presence
of Eve in the channel. Similarly, we have observed that for CR noise, a decoherence free subspace 
is formed by the quantum states
$\left\{\left|\psi^{+}\right\rangle,  \left|\phi^{-}\right\rangle\right\}$. Thus, if the communication channel contains both CD and CR noise, then the best choice for decoy qubits is the singlet state
$ \left|\phi^{-}\right\rangle$ state. Specifically, in such a situation we should form a verification string by applying PoP on an initial string formed as $ \left|\phi^{-}\right\rangle^{\otimes n}$. Interestingly, no such preferred state can be found for PD noise as we have observed that in PD noise, all the entangled states-based
subroutines show the same effect on the qubits traveling through the noisy
channel.

The present study also shows that in presence of CR noise, the BB84 subroutine would fail for
most of the values of rotation angle $\theta$  (specifically, for $\frac{\pi}{3}<\theta<\frac{2\pi}{3}$) as it yields negligibly small fidelity
at the receiver's end. A similar behavior is observed with $\left|\psi^{-}\right\rangle $
and $\left|\phi^{+}\right\rangle $ states for some other values
of rotation angle $\theta$. In brief, the present work provides a clear prescription on how to choose suitable decoy qubits for the preparation of verification string for performing a secure quantum task through a noisy channel whose characteristics are known. In some cases, specific types of noise may also be introduced intentionally to improve the security. We conclude the article with an expectation that experimentalists will find this work useful in designing environment-specific (depending upon which kind of noisy channel is present) schemes of secure quantum communication.

\end{document}